\documentclass[12pt,notoc,fleqn]{JHEP}

\usepackage{amssymb}
\usepackage{epsfig}
\usepackage{eucal}
\usepackage{amsmath}

\renewcommand{\cal}{\CMcal}                  
\newcommand{\ps}{\!\cdot\!}                  
\newcommand{\pa}{\mbox{\scriptsize\sf a}}    
\newcommand{\pb}{\mbox{\scriptsize\sf b}}    
\newcommand{\pq}{\mbox{\scriptsize\sf q}}    
\newcommand{\pg}{\mbox{\scriptsize\sf g}}    
\newcommand{\as}{\alpha_s}                   
\newcommand{\ab}{\overline{\alpha}_s}
\newcommand{\kk}{{\boldsymbol k}}            
\newcommand{\ku}{{\boldsymbol k}_1}
\newcommand{\kd}{{\boldsymbol k}_2}
\newcommand{\qq}{{\boldsymbol q}}
\newcommand{\dif}{{\rm d}}                   
\newcommand{\du}{\dif[\ku]}\newcommand{\dd}{\dif[\kd]}
\newcommand{\dk}{\dif[\kk]}

\newcommand{\dq}{\delta[\qq]}                
\newcommand{\G}{{\cal G}}                    
\newcommand{\K}{{\cal K}}                    
\newcommand{\M}{{\cal M}}                    
\renewcommand{\P}{{\mathcal P}}              
\renewcommand{\t}{{\boldsymbol t}}           
\newcommand{\e}{\varepsilon}                 
\renewcommand{\l}{\lambda}
\renewcommand{\o}{\omega}
\newcommand{\g}{\gamma}                   
\newcommand{\kp}{{\boldsymbol p}}         

\newcommand{\eq}[1]{eq.(\ref{#1})}         
\newcommand{\Eq}[1]{Eq.(\ref{#1})}         
\newcommand{\beq}{\begin{equation}}
\newcommand{\eeq}{\end{equation}}
\newcommand{\bea}{\begin{eqnarray}}
\newcommand{\eea}{\end{eqnarray}}
\newcommand{\non}{\nonumber}

\preprint{DFF/350/03/00 \\
TTP00-06}

\title{Heavy Quark Impact Factor at Next-to-leading Level
\thanks{Work supported in part by the E.U. QCDNET contract
FMRX-CT98-0194 and by MURST~(Italy).} }

\author{M. Ciafaloni \\ 
Dipartimento di Fisica, Universit\`a di Firenze
and INFN, Sezione di Firenze,
Largo E. Fermi 2, I-50125  Firenze, Italy.
Email: \email{ciafaloni@fi.infn.it}}

\author{G. Rodrigo \\
Institut f\"ur Teoretische Teilchenphysik,
Universit\"at Karlsruhe \\ D-76128 Karlsruhe, Germany.
Email: \email{rodrigo@particle.uni-karlsruhe.de}}

\abstract{
We further analyze the definition and the calculation of
the heavy quark impact factor at next-to-leading (NL)
$\log s$ level, and we provide its analytical expression
in a previously proposed $\kk$-factorization scheme. 
Our results indicate that $\kk$-factorization holds at NL level
with a properly chosen energy scale, and with the same gluonic
Green's function previously found in the massless probe case.}


\begin{document}

\section{Introduction}

Recent improvements~\cite{s98} of the next-to-leading 
$\log x$ (NL$x$) results~\cite{bfkl98}
in the BFKL framework, have stabilized the small-x 
behaviour in QCD, so that a phenomenological analysis 
of deep inelastic processes (DIS) seems now possible.

However, both the gluon density (satisfying the improved equation)
and the impact factors are needed in order to use 
$\kk$-factorization (Sec.~\ref{sec:kfactor})
to compute DIS or double DIS processes. So far, NL$x$ impact
factors have been found for the unphysical case of massless
initial quarks and gluons only~\cite{c98,impact99}. 
Partial features for massive quarks~\cite{fmassive99} 
and for colourless sources~\cite{fcolorless99} are known too.

In this paper we derive complete results for the case of initial 
massive quarks, with a twofold purpose. First, we want to check 
the validity of the $\kk$-factorization scheme introduced in 
Ref.~\cite{impact99}, or, in other words, to derive the same gluon 
Green's function with an explicit massive quark impact factor
which satisfies the expected collinear properties. 
Secondly, we develop as a byproduct some analytical 
techniques which are needed to deal with two-scale problems,
which are hopefully useful to cope with the physical cases also. 

The results of the paper rest on two observations. The first one, 
motivated in Sec.~\ref{sec:scale}, is that the factorized scale 
relevant in a high-energy two-scale process coupled to heavy quarks
is $s_0=Max(k_1,m_1)Max(k_2,m_2)$,
where $k_1$ and $k_2$ denote the relevant gluon virtualities,
rather than $s_0=k_1 k_2$, as in the massless quark case.
In fact, by subtracting the kernel contribution with such a scale
we are able in Sec.~\ref{sec:impact} to derive a result for the massive quark 
impact factor which is finite for $s\rightarrow \infty$, and has 
all the desired properties. 

The second observation is that we are able to disentangle the 
$(m/k)$-dependence of the impact factor by explicitly computing
its Mellin transform and its inverse. Given the singular 
energy dependence of the squared amplitude and of the phase 
space in the intermediate steps, this is by no means a trivial 
result and requires a careful handling of Mellin transform integrals in 
dimensional regularization, as explained in Sec.~\ref{sec:mellin}.

The outcome of such analysis is that the NL$x$ constant $H$-kernels,
previously introduced in the gluon Green's function~\cite{impact99},
are indeed probe-independent, and that the ensuing 
impact factors only contain factorizable single logarithmic 
collinear singularities. The use of such information in the 
improved small-$x$ equation and the left-over problems are 
discussed in Sec.~\ref{sec:conclusion}.

\section{k-Factorization in dijet production}

\label{sec:kfactor}

We consider the high-energy scattering of
two partons {\sf a} and {\sf b} with momenta $p_1$ and $p_2$
respectively.
Following~\cite{c98}, the colour averaged differential cross section
is factorized in a gauge-invariant way into a Green's function $\G_{\o}$
and impact factors $h_{\pa}$ and $h_{\pb}$ (Fig.~\ref{fig:fact})
\begin{equation}
 \frac{\dif\sigma_{\pa\pb}}{\du\,\dd}=
 \int\frac{\dif\o}{2\pi i\o}\,h_{\pa}(\ku)\G_{\o}(\ku,\kd)
 h_{\pb}(\kd)\left(\frac{s}{s_0(\ku,\kd)}\right)^{\o}~.
\label{fatt}
\end{equation}
We adopt $\dk=\dif^{2+2\e}\kk/\pi^{1+\e}$ as
transverse space measure.
The transverse plane is defined with respect to the incoming
momenta $p_1$ and $p_2$.
The transverse momenta  $\ku$ and $\kd$ play the role of
hard scales of the process. 
By definition, the impact factors are free of high-energy gluon
exchanges, which are subtracted out, but can still contain collinear
singularities which need to be factored out. 
The Green's function $\G_{\o}$ incorporates all the Regge-gluon
exchanges between the two partons.
The energy-scale $s_0(\ku,\kd)$ will be chosen later on.

At the next-to-leading $\log x$ (NL$x$) accuracy 
the Green's function $\G_{\o}$ has the following general form 
\begin{equation}
 \G_{\o}=(1+\ab H_L)\left[1-\frac{\ab}{\o}(K_0+K_{NL})\right]^{-1}\!
 (1+\ab H_R)~,
 \label{scomp}
\end{equation}
\FIGURE{\epsfig{file=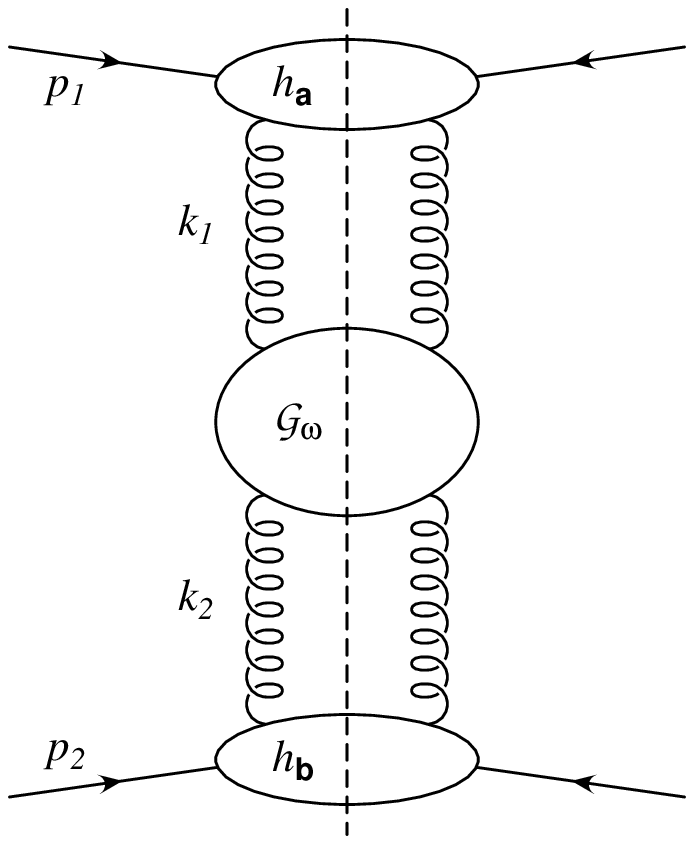,width=6cm}
\caption{Diagrammatic representation of $\kk$-factorization.}
\label{fig:fact}}
where $K_0$ and $K_{NL}$ are the leading $\log x$ (L$x$) and the  
NL$x$ BFKL kernels~\cite{bfkl98} respectively, 
$H_R (H_L)$ are operator factors 
introduced in~\cite{impact99} and 
\begin{equation*}
\ab = \frac{\as N_c}{\pi}~, \qquad
\as = \frac{g^2\Gamma(1-\e)\mu^{2\e}}{(4\pi)^{1+\e}}~,
\end{equation*}
is the dimensionless strong coupling constant.

As explained in~\cite{impact99}, the identification of the second
order impact factors, $h_{\pa}^{(1)}$ and $h_{\pb}^{(1)}$,
is affected by a double factorization scheme
ambiguity, due to both the choice of the scale $s_0$ and of the 
kernels $H_R (H_L)$.
The latter were introduced by Ciafaloni and Colferai (CC)
in~\cite{impact99} so as to provide 
partonic impact factors free of double $\log$ collinear 
divergences for the factorized scale choice $s_0=k_1 k_2$. 
It was also shown that the left-over single logarithmic divergences
could be factorized by the usual DGLAP approach.
A different factorization scheme, allowing double logarithmic 
divergences, was used instead in~\cite{fmassive99}, where an 
integral representation for the massive quark impact factor 
was presented also. 

In this paper we extend the CC scheme
to the massive quark case, by showing that the gluonic Green's function 
stays the same and that the collinear divergences of the massive impact
factor stay single logarithmic too.
Although the use of the $H$ kernels is optional for colourless
sources~\cite{fcolorless99}, for which they can be incorporated 
in the impact factors, we think that they help in stabilizing the 
collinear behaviour of the gluonic Green's function,
as already noticed in~\cite{c98}.

In the following we use the notation $p_1,p_2$ ($\l_1,\l_2$) for
the initial parton's momenta (helicities) and the indices $3,4$
(possibly $5$) for the final ones, with the Sudakov parametrization 
\begin{align*}
  k_1&=p_1-p_3\;
  =\;z_1 \bar{p}_1-\frac{\ku^2}{(1-z_1) s} \bar{p}_2+k_{1\perp}~, \\
  k_2&=p_2-p_4\;=
  \;-\frac{\kd^2}{(1-z_2) s}\bar{p}_1+z_2 \bar{p}_2+k_{2\perp}~,
\end{align*}
where we have introduced Sudakov variables $z_i$ and
transverse spacelike vectors $k_{i\perp}$ perpendicular to the
plane of the initial particle momenta light-cone basis
$\langle \bar{p}_1,\bar{p}_2\rangle$
\begin{equation*}
p_1 = \bar{p}_1 + \frac{m_1^2}{s} \bar{p}_2~, \qquad
p_2 = \bar{p}_2 + \frac{m_2^2}{s} \bar{p}_1~,
\end{equation*}
$\bar{p}_i\ps k_{j\perp}=0$, $\bar{p}_i^2=0$, $p_i^2=m_i^2$,
with $D-2$ Euclidean components $\kk_i:\kk_i^2=-k_{i\perp}^2>0$.
We also define $\qq=\ku+\kd$ as the transverse momentum of
parton~5. For simplicity, we use in the sequel
$q=|\qq|,k_i=|\kk_i|$ and we consider parton ${\sf b}$ as massless,
$m_2=0$, $m_1=m$.

\section{Factorization scheme and calculational procedure}

\FIGURE{\epsfig{file=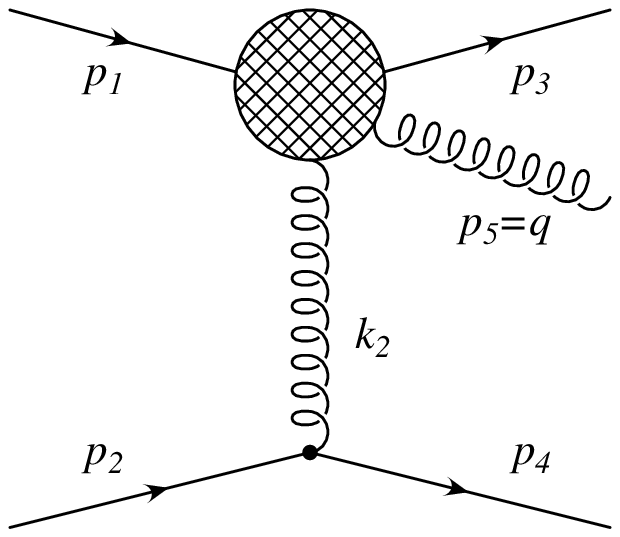,width=6cm}
\caption{Real gluon emission in the fragmentation region of quark ${\sf a}$.}
\label{fig:frag}}

Let's consider first the high-energy scattering of 
two partons ${\sf a}$ and
${\sf b}$ where ${\sf a}={\sf q}$ is a heavy quark of mass $m$
with real emission of an extra gluon ${\sf g}$ that we assume in the 
heavy quark fragmentation region (Fig.~\ref{fig:frag}).
In terms of invariants, we work in the kinematical region 
$s_2=(q+p_4)^2\simeq z_1 s \gg s_1=(p_3+q)^2\simeq q^2/z_1$,
so that $z_1>q/\sqrt{s}$ is the fragmentation phase space 
boundary.
The Born differential cross section
in this high energy region can be calculated in a straightforward
way (e.g., by eikonal coupling to the incoming parton ${\sf b}$),
and results to be
\begin{align}
& \frac{\dif\sigma_{\pq\pg\pb}}{\dif z_1\,\du\,\dd} =
 A_{\e} \, h_{\pb}^{(0)}(\kd)  \non \\ 
 \times \frac{1}{N_c} \bigg[ C_F \, \bigg(
  & \P_{\pg\pq}(z_1,\e) \,
 \frac{z_1^2}
 {\left[\qq^2+m^2z_1^2\right] \,
  \left[(\qq-z_1\kd)^2+m^2z_1^2\right]}   \non \\ 
 &  - \frac{m^2}{\kd^2} \,  
  \frac{z_1^3(1-z_1)\left[(2\qq-z_1\kd)\ps\kd\right]^2}
 {\left[\qq^2+m^2z_1^2\right]^2 \,
  \left[(\qq-z_1\kd)^2+m^2z_1^2\right]^2} \bigg)  \non \\ 
+   
  N_c \, \bigg( 
 & \P_{\pg\pq}(z_1,\e) \,
\frac{(1-z_1)\left[\qq\ps(\qq-z_1\kd)+m^2z_1^2\right]}
 {\left[\ku^2+m^2z_1^2\right] \, 
  \left[\qq^2+m^2z_1^2\right] \,
  \left[(\qq-z_1\kd)^2+m^2z_1^2\right]} \non \\ 
 & - \frac{m^2}{\kd^2} \,
 \frac{z_1(1-z_1)^2 \kd\ps(2\qq-\kd)\left[\kd\ps(2\qq-\kd)-z_1\kd^2\right]}
 {\left[\ku^2+m^2z_1^2\right]^2 \, 
  \left[\qq^2+m^2z_1^2\right] \,
  \left[(\qq-z_1\kd)^2+m^2z_1^2\right]} \bigg) \bigg]~,
\label{cross}
\end{align}
where
\begin{equation}
 \P_{\pg\pq}(z_1,\e)=\frac{1}{2z_1}\left[1+(1-z_1)^2+\e z_1^2\right]~,
 \label{pgq}
\end{equation}
is related to the quark to gluon splitting function,
\begin{equation}
 h_{}^{(0)}({\kk})=\sqrt{\frac{\pi}{N_c^2-1}}\;
 \frac{2C_F \as N_{\e}}{\kk^2\,\mu^{2\e}}~, \quad
 N_{\e} = \frac{(4\pi)^{\e/2}}{\Gamma(1-\e)}~,
\label{hzero}
\end{equation}
is the leading order impact factor, being the same for quarks and gluons,
$\mu$ is the renormalization scale, and 
\begin{equation}
A_{\e} =  \kk^2 \, h_{}^{(0)}({\kk}) \,
 \frac{\ab}{\Gamma(1-\e)\mu^{2\e}}~,
\end{equation}
is a constant that contains the dependence on the strong coupling
constant and some colour factors. 

Though complicated at first sight, \eq{cross} has some simple features 
that we now comment upon. First of all, only the $N_c$ part is 
really relevant to our purposes, the $C_F$ part being canceled with 
virtual correction upon $z_1$ and $\ku$ integration (see the 
following). We consider 
\begin{align}
& \frac{\dif\sigma_{\pq\pg\pb}}{\dif z_1\,\du\,\dd} \bigg|_{N_c} =
 A_{\e} \, h_{\pb}^{(0)}(\kd)  \non \\ 
 \times & \bigg[ \P_{\pg\pq}(z_1,\e) \,
\frac{(1-z_1)\left[\qq\ps(\qq-z_1\kd)+m^2z_1^2\right]}
 {\left[\ku^2+m^2z_1^2\right] \, 
  \left[\qq^2+m^2z_1^2\right] \,
  \left[(\qq-z_1\kd)^2+m^2z_1^2\right]} \non \\ 
 & - \frac{m^2}{\kd^2} \,
 \frac{z_1(1-z_1)^2 \kd\ps(2\qq-\kd)\left[\kd\ps(2\qq-\kd)-z_1\kd^2\right]}
 {\left[\ku^2+m^2z_1^2\right]^2 \, 
  \left[\qq^2+m^2z_1^2\right] \,
  \left[(\qq-z_1\kd)^2+m^2z_1^2\right]}  \bigg]~,
\label{crossNC}
\end{align}
The latter expression reduces, as expected to the known~\cite{impact99}
result for $m \rightarrow 0$, and matches the 
L$x$ differential cross section 
\begin{align}
& \frac{\dif\sigma^{(L)}_{\pq\pg\pb}}{\dif z_1\,\du\,\dd} =
h_{\pq}^{(0)}(\ku) \, h_{\pb}^{(0)}(\kd)  
\frac{\ab}{\qq^2\Gamma(1-\e)\mu^{2\e}} \, \frac{1}{z_1}~,
\label{Lxreal}
\end{align}
in the limit $z_1 \rightarrow 0$.
However, for~\eq{Lxreal} to be a good approximation to~\eq{crossNC},
we should require
\begin{equation}
z_1 \ll q/k_1~, \qquad z_1 \ll q/m~, \qquad z_1 \ll k_1/m~.
\label{constrains}
\end{equation}
The first two cutoffs can be summarized by $z_1 < q/Max(k_1,m)$, which
is a coherence condition for the case of heavy quarks, saying that 
the rapidity of the gluon cannot exceed that of the final quark. 
By integrating the leading expression~(\ref{Lxreal}) with the 
constraints~(\ref{constrains}) in the fragmentation region 
$z_1>q/\sqrt{s}$, we obtain
\begin{equation}
\frac{1}{h_{\pb}^{(0)}(\kd)} 
\frac{\dif\sigma^{(L)}_{\pq\pg\pb}}{\du\,\dd} =
\frac{\ab  h_{\pq}^{(0)}({\ku})}{\qq^2 \Gamma(1-\e)\mu^{2\e}}
\left( \log \frac{\sqrt{s}}{Max(k_1,m)} 
- \log \frac{q}{k_1} \Theta_{q k_1} \right)~.
\label{Lxintegrated}
\end{equation}
This expression is an estimate of the leading contribution contained 
in the complete result~\eq{cross}, which should be subtracted out 
in order to yield the impact factor in the massive quark case. 

\label{sec:scale}

Compared to the subtraction (or factorization) scheme adopted 
in~\cite{impact99} for 
$m=0$, the expression~(\ref{Lxintegrated}) differs 
by the replacement $k_1 \rightarrow Max(k_1,m)$, which leads,
by adding the symmetrical fragmentation region, to the choice 
for the factorized scale in~\eq{fatt}
\begin{equation}
s_0=Max(k_1,m_1)Max(k_2,m_2)~,
\label{scale}
\end{equation}
$m_1$ being the mass of quark ${\sf a}$
and $m_2$ the mass of quark ${\sf b}$.

By considering now both real and virtual contributions to the 
fragmentation function $F_{\pq}(z_1,\ku,\kd)$, we are led to introduce the 
following definition of the impact factor $h_{\pq}^{(1)}(\kk)$:
\begin{align}
& \int_{q/\sqrt{s}}^1 \dif z_1 \int \du 
F_{\pq}(z_1,\ku,\kd) \non \\
&= \int \du \, \ab \, h_{\pq}^{(0)}(\ku) \, K_0(\ku,\kd) \, 
\left( \log \frac{\sqrt{s}}{Max(k_1,m)} 
- \log \frac{q}{k_1} \Theta_{q k_1} \right) 
+ h_{\pq}^{(1)}(\kd)~,
\label{h1massive}
\end{align}
where
\begin{equation}
\ab \, K_0(\ku,\kd) = 
\frac{\ab}{\qq^2 \Gamma(1-\e)\mu^{2\e}}
+ 2 \o^{(1)}(\ku^2) \delta[\qq]~, \qquad 
\delta[\qq]=\pi^{1+\e} \delta^{2+2\e}(\qq)~,
\end{equation}
is the leading kernel, with 
\begin{equation}
 \o^{(1)}(\kk^2)=-\frac{g^2 N_c \kk^2}{(4\pi)^{2+\e}} \,
 \int\frac{\dif[{\boldsymbol p}]}{{\boldsymbol p}^2(\kk-{\boldsymbol p})^2}=
 -\frac{\ab}{2\e}\frac{\Gamma^2(1+\e)}{\Gamma(1+2\e)}
 \left(\frac{\kk^2}{\mu^2}\right)^{\e}~,
 \label{omega}
\end{equation}
the gluon Regge trajectory.
\Eq{h1massive} reduces for $m=0$, to the definition adopted 
in Ref.~\cite{impact99}, and in particular contains the 
subtraction term $\log q/k_1 \Theta_{q k_1}$ which
provides the expression ($H_R=H_L^{\dagger}=H$)
\begin{equation}
 H(\ku,\kd)=-\frac{1}{\qq^2\Gamma(1-\e)\mu^{2\e}}\,\log\frac{q}{k_1}\;
 \Theta_{q k_1}~, 
 \label{acca}
\end{equation}
for the $H$ kernel in the $\kk$-factorization formula. 

In order to simplify our subsequent calculations, we shall then use the 
known result~\cite{impact99} for $m=0$
\begin{align}
h_{\pq,m=0}^{(1)}(\kd)
&=h_{\pq}^{(0)}(\kd)\o^{(1)}(\kd^2)\left[\left(\frac{11}{6}-
 \frac{n_f}{3N_c}\right)+\left(\frac{3}{2}-\frac{1}{2}\e\right)
 \right. \non \\ & \left.
 - \left(\frac{67}{18}-\frac{\pi^2}{6}-\frac{5n_f}{9N_c}\right)\e \right]~,
\label{h1massless}
\end{align}
and we shall explicitly compute only the difference for a non 
vanishing mass. For this reason, we write the fragmentation vertex 
for a heavy quark as the massless fragmentation
vertex plus an extra quark mass dependent
contribution that cancels out for $m=0$
\begin{equation}
F_{\pq}(z_1,\ku,\kd) =
F^{m=0}_{\pq}(z_1,\ku,\kd) + \Delta F_{\pq}(z_1,\ku,\kd)~.
\label{splitF}
\end{equation}
Then, we find the following relationship 
between the massless quark impact factor and the heavy
quark impact factor
\begin{align}
h_{\pq}^{(1)}&(\kd) = h_{\pq,m=0}^{(1)}(\kd) 
+ \int_0^1 \dif z_1 \int \du \Delta F_{\pq}(z_1,\ku,\kd) \non \\
& + \int \du \, \ab \, h_{\pq}^{(0)}(\ku) \, K_0(\ku,\kd) \,
  \log \frac{m}{k_1} \, \Theta_{m \, k_1}~.
\label{h1mass}
\end{align}
The most complicated integral that remains in the r.h.s.
of \eq{h1mass} is the one for $\Delta F_{\pq}$ which will be 
calculated through its Mellin transform in $\kd$.
Notice also that the integration limits in $z_1$
have been extended down to $z_1=0$. Since $\Delta F_{\pq}$ is
regular at $z_1=0$ this change introduces only a negligible
error of order $1/s$. 

\section{Mellin transform and its inverse}

In order to perform the calculation outlined in~\eq{h1mass},
we proceed in two steps. First, we perform analytically the 
$\ku$ integration of~\eq{cross} by reducing the $\ku$-integrals
to two denominators, as explained in detail in Appendix~\ref{app:real}.
Then, we consider the virtual contributions~\cite{ffqk96}
quoted in Appendix~\ref{app:virtual}, and we organize them 
in terms of momentum fraction integrals only.
Finally, summing up real and virtual contributions to the fragmentation 
vertex (see \eq{real} and \eq{virtual} at the appendices)
(helicity non conserving not included)
we obtain the following expression for the difference $\Delta F_{\pq}(\kd)$,
arising from the second term in the r.h.s. of~\eq{h1mass}
\begin{align}
\Delta & F_{\pq}(\kd) = \Delta F_{\pq,real}(\kd)+\Delta F_{\pq,virt}(\kd)
 = A_{\e} \Bigg[
 \frac{\Gamma(-\e)}{2(1+2\e)} \frac{(m^2)^{\e}}{\kd^2} \non \\
& + \frac{\Gamma(1-\e)}{2} \bigg\{
\int_0^1 \int_0^1 \dif z_1 \, \dif x 
\left(\frac{1-z_1}{z_1} +\frac{1+\e}{2}z_1 \right) \non \\
& \quad \times \bigg[ \frac{1}{\left[x(1-x)\kd^2+m^2z_1^2\right]^{1-\e}}
-\frac{1}{\left[x(1-x)\kd^2\right]^{1-\e}} \bigg]  \non \\
& \quad + \frac{2m^2}{\kd^2} \int_0^1 \int_0^1
\frac{ z_1(1-z_1) \, \dif z_1 \, \dif x}
{\left[x(1-x)\kd^2+m^2z_1^2\right]^{1-\e}} \bigg\}  \Bigg]~,
\label{realvirtual}
\end{align}
where in $\Delta \Gamma^{(+)}_{\pq\pq}(\kd)$, \eq{gammapiu}
of Appendix~\ref{app:virtual},
the integration variable $x$ has been identified with
$z_1$ to simplify the sum.
Note again that, because of the subtraction of the $m=0$ part,
the $z_1$-integrals are convergent at $z_1=0$.

\subsection{Mellin integrals}

\label{sec:mellin}

To perform the last integrations in \eq{realvirtual} we 
first calculate its Mellin transform 
\begin{equation*}
\Delta \tilde{F}_{\pq}(\gamma) =  
\Gamma(1+\e) \, (m^2)^{-\e} \int \dd  \left( \frac{\kd^2}{m^2}\right)^{\g-1}
\Delta F_{\pq}(\kd)~,  
\end{equation*}
yielding
\begin{align}
\Delta&\tilde{F}_{\pq}(\gamma) = A_{\e} \, (m^2)^{\e} 
 \frac{\Gamma(\g+\e)\Gamma(1-\g-2\e)\Gamma^2(1-\g-\e)}
 {8\Gamma(2-2\g-2\e)} \non \\
 \times & 
\bigg[ \frac{1+\e}{\g+2\e} + \frac{2}{1-2\g-4\e}
\left( \frac{1}{1-\g-2\e}- \frac{1}{3-2\g-2\e} \right) \bigg]~.
\label{mellin}
\end{align}
Although this Mellin transform is finite for $\e \rightarrow 0$
the limit $\e=0$ cannot be taken in this expression.
In fact, it is straightforward, though not trivial, to show 
that the r.h.s. of~\eq{realvirtual} behaves as
\begin{equation}
\Delta F_{\pq}(\kd) \underset{k2\ll m}{\simeq} (\kd^2)^{\e-1}~, \qquad
\Delta F_{\pq}(\kd) \underset{k2\gg m}{\simeq} (\kd^2)^{-1} (m^2)^{\e}~.
\end{equation}
Therefore, the Mellin transform converges only in the small band
$1-2\e < Re \g <1-\e$ and the $\e$ dependence should 
be kept until the end. 

To recover $\Delta F_{\pq}(\kd)$
we should calculate the following inverse Mellin transform
\begin{equation*}
\Delta F_{\pq}(\kd) =
\frac{1}{m^2} \int_{1-2\e < Re \g <1-\e} \frac{\dif \g}{2\pi i}  
\left(\frac{\kd^2}{m^2}\right)^{-\g-\e}
\Delta \tilde{F}_{\pq}(\gamma)~.  
\end{equation*}
We consider first the limit $\kd^2>m^2$. Then, we displace the
integration contour around the positive real semiaxis, enclosing
all the poles placed in $\g \ge 1-\e$, the smaller one giving the
smaller power in $m/k_2$. The first pole, at $\g=1-\e$,
yields the following result
\begin{align}
\Delta F_{\pq}(\kd)&= \ab \, h_{\pq}^{(0)}(\kd) \,
\bigg\{  - \frac{2+3\e+2\e^2}{4\e^2(1+2\e)}
\left(\frac{m^2}{\mu^2}\right)^{\e} 
+\cal{O}(m/k_2) \bigg\}~. 
\label{DFm}
\end{align}
For $\kd^2<m^2$ we displace the integration contour around
the negative real semiaxis, enclosing all the poles placed in
$\g \le 1-2\e$. As before, the first pole, at $\g=1-2\e$, gives the
answer at order $\cal{O}(k_2/m)$
\begin{align}
\Delta F_{\pq}(\kd)&= h_{\pq}^{(0)}(\kd) \, \bigg\{ 
\o^{(1)}(\kd^2) \bigg[ - \frac{1+5\e-2\e^2}{2(1+2\e)} 
- \log \left(\frac{\kd^2}{m^2}\right) \non \\
& +\psi(1-\e)-\psi(1)-2\psi(\e)+2\psi(2\e) \bigg]
+\cal{O}(k_2/m) \bigg\}~.
\label{DFk}
\end{align}
Since $\o^{(1)} \sim 1/\e$ is infrared singular (cf. \eq{omega}),
both formulas show double logarithmic singularities of type
$1/\e^2$ and $1/\e \log(\kd^2/m^2)$.

The last ingredient we need in order to extract the next-to-leading
quark impact factor is the last term in the r.h.s. of~\eq{h1mass}.
For the real emission contribution to $K_0$ we get the integral
\begin{equation}
  I_m = \int \du \frac{\ab  h_{\pq}^{(0)}({\ku})}{\qq^2 \Gamma(1-\e)\mu^{2\e}}
  \log \frac{m}{k_1} \, \Theta_{m \, k_1}~. 
\end{equation}
We use the following representation
\begin{equation*}
 \log\frac{a}{b} \,\Theta_{ab}=\lim_{\alpha\to0^+}
 \int_{-i\infty}^{+i\infty}\frac{\dif\l}{2\pi i}\,\frac{1}{(\l+\alpha)^2}
 \left(\frac{a}{b}\right)^{\l}
 \equiv \int \dif[\l] \, \left(\frac{a}{b}\right)^{\l}~,
\end{equation*}
valid for $a,b>0$, which allows us to write
\begin{align}
I_m & = \frac{A_{\e}}{2}
\int \dif[\l] \, (m^2)^{\l} \int \frac{\du}
{\qq^2 \, (\ku^2)^{1+\l}}  \non \\
& = \frac{A_{\e}}{2} \int \dif[\l]   
\frac{\Gamma(1+\l-\e) \Gamma(\e) \Gamma(\e-\l)}
{\Gamma(1+\l) \Gamma(2\e-\l)} (m^2)^{\l} (\kd^2)^{-1-\l+\e}~.
\label{Im}
\end{align}
The integrand vanishes for $|\l|\rightarrow \infty$ in all directions
apart from the real axis. As before, we consider first the case
$\kd^2>m^2$ and displace the integration contour around the
positive real semiaxis enclosing all the poles placed in $\l>0$.
The smaller pole, at $\l=\e$, gives us the result at order
$\cal{O}(m/k_2)$ 
\begin{align}
I_m &= \ab \, h_{\pq}^{(0)}(\kd) \bigg\{   
\frac{1}{2\e^2\Gamma(1-\e)\Gamma(1+\e)}
\left(\frac{m^2}{\mu^2}\right)^{\e} 
+ \cal{O}(m/k_2) \bigg\}~. 
\label{Im1}
\end{align}
On the other hand, for $\kd^2<m^2$, we consider the poles
placed at the negative real semiaxis and therefore
the residue at $\l=-\alpha$ with $\alpha \rightarrow 0^+$.
By taking into account also the virtual contribution to $K_0$ in 
this case, we obtain 
\begin{align}
I_m &- h_{\pq}^{(0)}(\kd) \, \o^{(1)}(\kd^2) 
\log \left(\frac{\kd^2}{m^2}\right) =
h_{\pq}^{(0)}(\kd) \bigg\{ \o^{(1)}(\kd^2) 
\bigg[ \log \left(\frac{\kd^2}{m^2}\right) \non \\
& +2 \left[ \psi(1)-\psi(1-\e)+\psi(\e)-\psi(2\e) \right] \bigg]
+ \cal{O}(k_2/m)\bigg\}~.
\label{Im2}
\end{align}
\Eq{Im1} and \eq{Im2} show double $\log$ singularities also. 

\subsection{Impact factors}

\label{sec:impact}

Finally, summing up all the pieces according to~\eq{h1mass}, the 
impact factor for heavy quarks at the next-to-leading
level can be written as 
\begin{equation}
h_{\pq}(\kd) = h_{\pq}^{(0)}(\kd) + h_{\pq}^{(1)}(\kd)~,      
\end{equation}
where
\begin{align}
& h_{\pq}^{(1)}(\kd)  
 = h_{\pq,m=0}^{(1)}(\kd) + h_{\pq}^{(0)}(\kd) \non \\
\times & \bigg\{ \o^{(1)}(m^2) \frac{\Gamma(1+2\e)}{\e \Gamma^2(1+\e)}
\bigg[ \frac{2+3\e+2\e^2}{2(1+2\e)} 
- \frac{1}{\Gamma(1-\e)\Gamma(1+\e)} \bigg]
+ \cal{O}(m/k_2) \bigg\}~, 
\label{h1m}
\end{align}
is valid in the limit $\kd^2>m^2$, and
\begin{align}
& h_{\pq}^{(1)}(\kd)  
 = h_{\pq,m=0}^{(1)}(\kd) + h_{\pq}^{(0)}(\kd)  \non \\
\times & \bigg\{ \o^{(1)}(\kd^2) 
\bigg[ \psi(1) - \psi(1-\e) - \frac{1+5\e-2\e^2}{2(1+2\e)} \bigg] 
+\cal{O}(k_2/m) \bigg\}~,
\label{h1k}
\end{align}
is valid for $\kd^2<m^2$.

We notice, in the first place, that all double $\log$ contributions
of type $1/\e^2$ and $1/\e \log(\kd^2/m^2)$ appearing in 
eqs.~(\ref{DFm}-\ref{DFk}) and (\ref{Im1}-\ref{Im2}) have canceled 
out in eqs.~(\ref{h1m}) and~(\ref{h1k}).
This means that indeed our subtraction of the leading kernel was
effective, thus lending credit to the scale~(\ref{scale}) and to the 
$H$-kernel~(\ref{acca}).

The remaining singularities of the impact factor are single 
logarithmic ones~$\sim~1/\e$, having the structure
\begin{align}
\left. h_{\pq}^{(1)}(\kd) \right|_{sing} &=
h_{\pq}^{(0)}(\kd)  \bigg(
\frac{3}{2} \o^{(1)}(\kd^2)
-\frac{1}{2} \o^{(1)}(m^2) \Theta_{k_2 \, m}
-\frac{1}{2} \o^{(1)}(\kd^2) \Theta_{m \, k_2} \bigg)~.
\label{h1sing}
\end{align}
Here the first term has the customary collinear 
interpretation~\cite{impact99}, as coming from the finite part
of the ${\sf q} \rightarrow {\sf g}$ anomalous dimension
\begin{equation}
\gamma_{\pg \pq} - \frac{C_F \as}{\pi \omega} = 
- \frac{C_F \as}{2\pi} \left(\frac{3}{2}-\frac{1}{2}\e \right)~,
\end{equation}
while the remaining ones -- depending on the scale $Min(k_2,m)$ --
do not have such interpretation. Note, however, that a finite 
mass scale change $m\rightarrow c \, m$, will produce exactly this 
type of contributions from the singular integration of the $K_0$
kernel in~\eq{h1massive} acting on $h_{\pq}^{(0)}(\ku)$ over the
region $0<k_1<Min(k_2,m)$, leading to the expression 
\begin{align}
\delta h_{a}^{(1)}(\kd) = 
h_{\pq}^{(0)}(\kd) \,  a(c) \bigg[
 \o^{(1)}(m^2) \Theta_{k_2 \, m}
+ \o^{(1)}(\kd^2) \Theta_{m \, k_2} \bigg]~.
\label{deltah}
\end{align}
Therefore, the singularities (\ref{h1sing}) can finally be interpreted 
in the form 
\begin{align}
\left. h_{\pq}^{(1)}(\kd) \right|_{sing} 
&= h_{\pq}^{(0)}(\kd) \, 
\frac{3}{2} \left[ 
\o^{(1)}(\kd^2) - \o^{(1)}(m^2) \right] \Theta_{k_2 \, m}
+ \delta h_{1}^{(1)}(\kd) \non \\
&= h_{\pq}^{(0)}(\kd) \, \frac{\as N_c}{2\pi}
\left( - \frac{3}{2} \log \frac{\kd^2}{m^2} \right) \Theta_{k_2 \, m}
+ \delta h_{1}^{(1)}(\kd)~,
\label{h1singular}
\end{align}
meaning that $h_{\pq}^{(1)}$ is actually {\it finite}, with the 
$\log(\kd^2/m^2)$ dependence predicted by the DGLAP equations,
apart from a proper mass scale change in~\eq{acca}. In other words, the
scale leading to a finite massive quark impact factor differs 
from~\eq{scale} by a finite renormalization of the quark mass, 
which is a normal ambiguity in this type of problems. 

Our final result for the heavy quark impact factor 
at the next-to-leading level reads
\begin{align}
& h_{\pq}(\kd) = \left. h_{\pq}^{(1)}(\kd) \right|_{sing} +
\left. h_{\pq}(\kd) \right|_{finite}~,
\end{align}
where the singular piece is defined in~\eq{h1singular} and 
\begin{align}
\left. h_{\pq}(\kd) \right|_{finite}
&= h_{\pq}^{(0)}(\as(\kd)) \non \\
\times & \bigg\{ 1+\frac{\as N_c}{2\pi} \bigg[
\K - \frac{\pi^2}{6} - \bigg(\frac{3}{2} + 
\sum_{Re\g >1} Res [\tilde{h}(\g)] \bigg) \Theta_{k_2 \, m}
\non \\
& + \bigg( 2 +  
\sum_{Re\g <1} Res [\tilde{h}(\g)] \bigg) \Theta_{m \, k_2}
\bigg] \bigg\}~,
\label{final}
\end{align}
is the finite contribution to the heavy quark impact factor, with
\begin{equation}
\K = \frac{67}{18}-\frac{\pi^2}{6}-\frac{5n_f}{9N_c}~,
\end{equation}
the constant term of the impact factor for massless quarks,
\eq{h1massless}, and 
\begin{align}
\tilde{h}(\g) &= 
\left( \frac{\kd^2}{m^2}\right)^{1-\g} \bigg\{
 \frac{\Gamma(\g)\Gamma^3(1-\g)}
 {4\Gamma(2-2\g)}
\bigg[ \frac{1}{\g} + \frac{2}{1-2\g}
\left( \frac{1}{1-\g}- \frac{1}{3-2\g} \right) \bigg] \non \\
& - \frac{1}{(1-\g)^2}
\left[ \psi(1-\g)+\psi(\g)-2\psi(1) \right]
\bigg\}~.
\label{hgamma}
\end{align}
The compact expression~(\ref{hgamma}) was obtained
by adding to the Mellin transform~(\ref{mellin})
the contribution of~\eq{Im}, with $\l \rightarrow -1+\g+2\e$,
and the Mellin transform of the virtual piece of the
last term at the r.h.s of~\eq{h1mass}.
The sum of these three terms, apart from
$\g = 1-\e,1-2\e$, whose contributions have already been 
treated separately, is finite
and therefore was expanded for $\e \rightarrow 0$.
As for the massless case, the singularities proportional to
$(11/6-n_f/3N_c)$, the beta function, were absorbed by the running
strong coupling constant $\as(\kd)$.
The function $\tilde{h}(\g)$
provides the corrections of order $\cal{O}(m/k_2)$ and $\cal{O}(k_2/m)$
to the impact factor for $m^2<\kd^2$ and $m^2>\kd^2$ 
respectively, yielding the following final result
\begin{align}
\sum_{Re\g >1} & Res [\tilde{h}(\g)] = Li_2\left(\frac{m^2}{\kd^2}\right)
+\sum_{n=1}^{\infty} \frac{\Gamma(2n+2)}{\Gamma^2(n+1)}
\left(-\frac{m^2}{\kd^2}\right)^n \non \\
\times & \bigg\{
\frac{1}{n^2}-\frac{1}{2(n+1)^2}-\frac{1}{2(2n-1)^2}
-\frac{6}{(2n+1)^3}-\frac{3}{2(2n+1)^2} \non \\
& + \bigg( \frac{2}{n}-\frac{1}{n+1}-\frac{1}{2(2n-1)}
-\frac{3}{(2n+1)^2}-\frac{3}{2(2n+1)} \bigg) \non \\
& \times \bigg(\psi(n+1) - \psi(2n+2) 
- \frac{1}{2}\log \frac{m^2}{\kd^2}\bigg)
\bigg\}~,
\end{align}
and
\begin{align}
 \sum_{Re\g <1} & Res [\tilde{h}(\g)] = Li_2\left(\frac{\kd^2}{m^2}\right)
 -\frac{3\pi^2}{8} \sqrt{\frac{\kd^2}{m^2}}
+ \frac{\kd^2}{m^2}
\left(\frac{5}{6}-\frac{1}{4} \log \frac{\kd^2}{m^2} \right) \non \\ &
+ \sum_{n=1}^{\infty} 
\frac{\Gamma^2(n+1)}{\Gamma(2n+2)} \left(-\frac{\kd^2}{m^2}\right)^{n+1} 
\left(\frac{1}{n}+\frac{2}{n+1}-\frac{3}{2n+1}-\frac{1}{2n+3}\right)~.
\end{align}

\section{Conclusions}

\label{sec:conclusion}

Starting from the explicit squared matrix element for gluon
emission in~\eq{cross} we have motivated the subtraction of the
leading term in~\eq{Lxintegrated}, and we have performed the $\ku$
and $z_1$ integrals needed to provide an explicit result for the 
heavy quark impact factor in~\eq{h1singular} and~\eq{final}.

Even if the cross section being investigated is unphysical,
the relevance of our results stems from the consistency of the
following features: (i) the validity of the $\kk$-factorization
formula~(\ref{fatt}) with scale $s_0=Max(k_1,m_1)Max(k_2,m_2)$;
(ii) the explicit expression of the impact factor with
factorizable single logarithmic collinear divergences, and
(iii) the probe-independence of the subleading $H$-kernels 
of the CC scheme~\cite{impact99}, defined in \eq{acca}.

Even if such universal extra kernels can be reabsorbed in the
impact factors for colourless sources~\cite{fcolorless99},
they help clarifying the structure of the collinear limits
for two-scale processes as elaborated at length by CCS~\cite{s98}.
Here, it was shown that the gluonic Green's function including
such kernels, is free of double $\log$s of collinear type
for both $k_1 \gg k_2$ and $k_2 \gg k_1$. As a consequence, even 
colourless impact factors will show, in the present scheme,
simple collinear properties, as expected from their DGLAP
analysis~\cite{taiuti}.

Of course, the real problem is to provide an
explicit expression for the DIS impact factors. But -- if the
lesson learned form the L and NL kernels is still valid --
the impact factor's magnitude is not expected to be much 
different from their approximate collinear evaluation.

\section*{Aknowledgements}

We wish to thank Gianni Camici for very valuable contributions in the
early stages of this paper. 
Work supported in part by the E.U. QCDNET contract
FMRX-CT98-0194 and by MURST (Italy). G.R. acknowledges financial 
support from INFN (Italy) and BMBF Project 05HT9VKB0 (Germany).

\appendix

\section{Real contribution to the fragmentation vertex}

\label{app:real}

The differential cross section for real gluon emission off a heavy
quark, \eq{cross}, can be simplified by using identities of the 
following type
\begin{align}
& \frac{z_1^3(1-z_1)\left[(2\qq-z_1\kd)\ps\kd\right]^2}
 {\left[\qq^2+m^2z_1^2\right]^2 \,
  \left[(\qq-z_1\kd)^2+m^2z_1^2\right]^2} = z_1(1-z_1) \left\{
  \frac{1}{\left[\qq^2+m^2z_1^2\right]^2} \right. \non \\
& \left. 
 + \frac{1}{\left[(\qq-z_1\kd)^2+m^2z_1^2\right]^2}  
 - \frac{2}{\left[\qq^2+m^2z_1^2\right] \,
   \left[(\qq-z_1\kd)^2+m^2z_1^2\right]} \right\}~,
\end{align}
to split the full expression into several contributions 
with at most two different propagators free of $\qq$
dependences at the numerator. 
After some algebra, we  obtain the following expression for
the real contribution to the fragmentation vertex of quark ${\sf q}$
\begin{align}
 F_{\pq,real}&(z_1,\ku,\kd)
 = A_{\e} \bigg\{
- z_1 (1-z_1) \frac{m^2}{\kd^2} \bigg[ \frac{C_F}{N_c}
\bigg( \frac{1}{\left[\qq^2+m^2z_1^2\right]^2}  \non \\ 
& + \frac{1}{\left[(\qq-z_1\kd)^2+m^2z_1^2\right]^2} \bigg) 
+ \frac{1}{\left[\ku^2+m^2 z_1^2\right]^2} \bigg] \non \\
& + \left(\frac{C_F}{N_c}-\frac{1}{2}\right)
\frac{z_1^2 \P_{\pg\pq}(z_1,\e) + 2 z_1(1-z_1)(m^2/\kd^2)  }
{\left[\qq^2+m^2z_1^2\right] \,
 \left[(\qq-z_1\kd)^2+m^2z_1^2\right]} \non \\
& + \frac{\P_{\pg\pq}(z_1,\e) + 2 z_1(1-z_1)(m^2/\kd^2)}
 {2 \left[\ku^2+m^2z_1^2\right] \, 
  \left[\qq^2+m^2z_1^2\right] } \non \\
& + \frac{(1-z_1)^2\P_{\pg\pq}(z_1,\e) + 2 z_1(1-z_1)(m^2/\kd^2)}
 {2 \left[\ku^2+m^2z_1^2\right] \, 
  \left[(\qq-z_1\kd)^2+m^2z_1^2\right]}  \bigg\}~.
\end{align}
According to~\eq{h1massive}, this expression has to be 
integrated first in $\ku$ and then in $z_1$, for $z_1>q/\sqrt{s}$.
To perform the first integration we use 
\begin{align}
& \int \frac{\du}
{\left[\ku^2+m^2z_1^2\right]^\alpha \, 
\left[(\ku+\kp)^2+m^2z_1^2\right]^\beta} = \non \\
& \qquad \qquad
  \frac{\Gamma(\alpha+\beta-1-\e)}{\Gamma(\alpha)\Gamma(\beta)}
  \int_0^1 dx \frac{x^{\alpha-1}(1-x)^{\beta-1}}
  {\left[x(1-x)p^2+m^2z_1^2\right]^{\alpha+\beta-1-\e}}~,
\end{align}
then, we obtain
\begin{align}
& F_{\pq,real}(z_1,\kd) =
\int \du  F_{\pq,real}(z_1,\ku,\kd) \non \\ 
& = A_{\e} \Gamma(1-\e) \bigg\{
- \left(\frac{2C_F}{N_c}+1\right)
z_1^{-1+2\e} (1-z_1)\frac{(m^2)^{\e}}{\kd^2} \non \\
& + \int_0^1 \dif x \, \bigg[
\left(\frac{C_F}{N_c}-\frac{1}{2}\right)
\frac{z_1^{2\e} \P_{\pg\pq}(z_1,\e) + 2 z_1^{-1+2\e}(1-z_1)(m^2/\kd^2)  }
{\left[x(1-x)\kd^2+m^2\right]^{1-\e} } \non \\
& + \frac{\P_{\pg\pq}(z_1,\e) + 2 z_1(1-z_1)(m^2/\kd^2)}
 {2 \left[x(1-x)\kd^2+m^2z_1^2\right]^{1-\e} } \non \\
& + \frac{(1-z_1)^2\P_{\pg\pq}(z_1,\e) + 2 z_1(1-z_1)(m^2/\kd^2)}
 {2 \left[x(1-x)(1-z_1)^2\kd^2+m^2z_1^2\right]^{1-\e} } \bigg] \bigg\}~.
\end{align}
By subtracting the massless contribution 
\begin{equation}
\Delta F_{\pq,real}(z_1,\kd) = 
F_{\pq,real}(z_1,\kd)-F^{m=0}_{\pq,real}(z_1,\kd)~,
\end{equation}
we get an expression that is regular at $z_1=0$
and therefore can be integrated 
down to $z_1=0$ without changing the final result. 
Let's define
\begin{align}
\Delta& F_{\pq,real}(\kd) = 
 \int_0^1 \dif z_1 \, \Delta F_{\pq,real}(z_1,\kd) \non \\
 & = A_{\e}  \Bigg[
\left(\frac{C_F}{N_c}+\frac{1}{2}\right)
 \frac{\Gamma(-\e)}{1+2\e} \frac{(m^2)^{\e}}{\kd^2} \non \\
& - \left(\frac{C_F}{N_c}-\frac{1}{2}\right) \frac{\Gamma(-\e)}{2} 
\bigg\{ \left( \frac{1}{1+2\e}+\frac{\e}{2} \right) \non \\
& \quad \times  \int_0^1 \dif x \bigg[
 \frac{1}{\left[x(1-x)\kd^2+m^2\right]^{1-\e}} 
 -\frac{1}{\left[x(1-x)\kd^2\right]^{1-\e}} \bigg] \non \\ 
& \quad + \frac{2(m^2/\kd^2)}{1+2\e}
\int_0^1 \frac{\dif x}{\left[x(1-x)\kd^2+m^2\right]^{1-\e}} \bigg\}\non \\
& + \frac{\Gamma(1-\e)}{2} \bigg\{
\int_0^1 \int_0^1 \dif z_1 \, \dif x 
\left(\frac{1-z_1}{z_1} +\frac{1+\e}{2}z_1 \right) \non \\
& \quad \times \bigg[ \frac{1}{\left[x(1-x)\kd^2+m^2z_1^2\right]^{1-\e}}
-\frac{1}{\left[x(1-x)\kd^2\right]^{1-\e}}  \non \\
& \quad + \frac{(1-z_1)^2}{\left[x(1-x)(1-z_1)^2\kd^2+m^2z_1^2\right]^{1-\e}}
- \frac{(1-z_1)^2}{\left[x(1-x)(1-z_1)^2\kd^2\right]^{1-\e}} \bigg] \non \\
& \quad + \frac{2m^2}{\kd^2}
\int_0^1 \int_0^1 z_1(1-z_1) \, \dif z_1 \, \dif x \bigg[
\frac{1}{\left[x(1-x)\kd^2+m^2z_1^2\right]^{1-\e}} \non \\
& \qquad + \frac{1}{\left[x(1-x)(1-z_1)^2\kd^2+m^2z_1^2\right]^{1-\e}} 
\bigg] \bigg\}  \Bigg]~.
\label{real}
\end{align}
Notice that some of the integrations has been kept undone.
This long expression, although cumbersome, will be drastically simplified 
after adding the virtual contribution before doing any further 
integration.  

\section{Virtual contribution to the fragmentation vertex}

\label{app:virtual}

The correction to the cross section due to virtual emission, including
subleading effects, for general parton-parton scattering, can be
extracted from the amplitude of Ref.~\cite{ffqk96} ($t=-\ku^2$)
\begin{align}
 \M_{\pa\pb} &= 2s g^2 (\t_{\pa}^c \t_{\pb}^c)
 \Big[
 \delta_{\l_3,\l_1} \Big(1+\Gamma^{(+)}_{\pa\pa}\Big)
 + \delta_{\l_3,-\l_1} \Gamma^{(-)}_{\pa\pa} \Big]
 \frac{1}{t}
 \left[1+\o(-t)\log\frac{s}{-t}\right] \non \\
& \times \Big[ 
 \Big(1+\Gamma^{(+)}_{\pb\pb}\Big) \delta_{\l_4,\l_2}
 + \Gamma^{(-)}_{\pb\pb} \delta_{\l_4,-\l_2} \Big]~,
\label{corrvirt}
\end{align}
were $\Gamma^{(+)}$ and $\Gamma^{(-)}$ are the helicity conserving 
and the helicity non-conserving contributions respectively. 
At the order we are working, the virtual terms contribute to the 
fragmentation vertex as follows
\begin{equation}
 F_{\pq,virt}(z_1,\ku,\kd) = 
 h^{(0)}_{\pq}(\ku)\,2\Gamma^{(+)}_{\pq\pq}(\ku)\delta(1-z_1)\dq~.
\end{equation}
As for real emission, we split $\Gamma_{\pq\pq}^{(+)}$ into  
a massless contribution plus a quark mass dependent extra term 
\begin{equation}
\Gamma_{\pq\pq}^{(+)} = \Gamma_{\pq\pq,m=0}^{(+)}+ \Delta
\Gamma_{\pq\pq}^{(+)}~,
\end{equation}
where $\Delta \Gamma_{\pq\pq}^{(+)}$ cancels for $m=0$, being
\begin{align}
\Gamma_{\pq\pq,m=0}^{(+)}(\kk) & = \frac{\o^{(1)}(\kk^2)}{2} \bigg\{  
\psi(1-\e) - 2 \psi(\e) + \psi(1) \non \\
& + \frac{1}{1+2\e}\left(\frac{1}{4(3+2\e)}-\frac{2}{\e}-\frac{7}{4}
-\frac{n_f}{N_c} \frac{1+\e}{3+2\e} \right) - \frac{1}{2} \non \\
& + \frac{2}{\e} \, \frac{C_F}{N_c} 
\left(\frac{1}{1+2\e}+\frac{\e}{2} \right) 
\bigg\}~,
\end{align}
and
\begin{align}
\Delta \Gamma_{\pq\pq}^{(+)}&(\kk) = 
\frac{\ab}{4\Gamma(1-\e)\mu^{2\e}} \Bigg[
\Gamma(-\e) \left(\frac{C_F}{N_c}-\frac{1}{2}\right) \bigg\{ 
\left(\frac{1}{1+2\e} + \frac{\e}{2} \right) \non \\
& \times \int_0^1 \dif x  
\bigg[ \frac{1}{\left[ x(1-x)\kk^2+m^2\right]^{1-\e}}
-\frac{1}{\left[ x(1-x)\kk^2\right]^{1-\e}} \bigg] \non \\
& + \frac{2(m^2/\kk^2)}{1+2\e} \int_0^1 
\frac{\dif x}{\left[ x(1-x)\kk^2+m^2\right]^{1-\e}} \bigg\}
- \frac{C_F}{N_c} \frac{2 \Gamma(-\e)}{1+2\e}\frac{(m^2)^{\e}}{\kk^2} \non \\
& - \Gamma(1-\e) \bigg\{ \int_0^1 \int_0^1  \dif x \,  \dif y  
(1-x)^2 \left(\frac{1-x}{x}+\frac{1+\e}{2} x \right) \non \\
& \times \bigg[ \frac{1}{\left[y(1-y)(1-x)^2 \kk^2+m^2 x^2\right]^{1-\e}}
-\frac{1}{\left[y(1-y)(1-x)^2 \kk^2\right]^{1-\e}} \bigg] \non \\
& + \frac{2m^2}{\kk^2} \int_0^1 \int_0^1  \frac{x(1-x) \dif x \, \dif y}
{\left[y(1-y)(1-x)^2 \kk^2+m^2 x^2\right]^{1-\e}}
\bigg\} \Bigg]~.
\label{gammapiu}
\end{align}

We define 
\begin{align}
 \Delta F_{\pq,virt}(\kd) &= \int_0^1 \dif z_1 \int \du
 \Delta F_{\pq,virt}(z_1,\ku,\kd) \non \\
&= h^{(0)}_{\pq}(\kd)\,2\Delta \Gamma^{(+)}_{\pq\pq}(\kd)~.
\label{virtual}
\end{align}
Adding up together the real, \eq{real}, and the virtual, \eq{virtual},
contributions to $\Delta F_{\pq}(\kd)$ most of the terms cancel.
To see this cancellation it is enough to identify 
the integration variable $x$ appearing in 
$\Delta \Gamma^{(+)}_{\pq\pq}(\kd)$, \eq{gammapiu}, with 
the momentum fraction $z_1$ appearing in \eq{real}.
In particular, the $C_F$ contribution fully cancels and the $N_c$
part gives the the simplified result presented in \eq{realvirtual}.


\end{document}